\documentclass[twocolumn,reprint,amsmath,amssymb,aps]{revtex4-2}
\usepackage{graphicx}
\usepackage{dcolumn}
\usepackage{bm}
\usepackage{float}
\usepackage{tikz}

\usepackage{quantikz}

\begin{document}

\title{Pulsed coherent and squeezed driving of a spin-$S$ emitter as a deterministic source of Wigner-negative light}

\author{Rory Robertson}
\email{rory.robertson@auckland.ac.nz}
\author{Scott Parkins}
\email{s.parkins@auckland.ac.nz}
\affiliation{Dodd-Walls Centre for Photonic and Quantum Technologies, New Zealand}
\affiliation{Department of Physics, University of Auckland, Auckland 1010, New Zealand}

\date{September 21, 2026}

\begin{abstract}
    Generating propagating Wigner-negative states of light is an important and difficult task for many quantum technologies. Historically, the proposed methods of creating such states have been intrinsically probabilistic, relying on conditional measurements or photon heralding. More recently, methods have been proposed involving the steady-state driving of two-level emitters, where Wigner-negative temporal modes are present in the outgoing field. We theoretically demonstrate through the simulation of a cascaded-systems model that this approach can be improved significantly with pulsed coherent or squeezed driving of a spin-1/2 emitter. This enables the deterministic generation of strongly Wigner-negative states, using only driving-source states with positive Gaussian Wigner distributions. This method is extended to the driving of a spin-$S$ emitter with $S\geq 1$ to generate still more exotic, highly non-classical states. The array of states demonstrated in this work includes close approximations to displaced $N$-photon states, squeezed Schr\"odinger cat states, and finite-energy Gottesman-Kitaev-Preskill (GKP) states.
    
\end{abstract}

\maketitle

\section{Introduction}\label{section:introduction}
The generation of propagating modes of light displaying negativity in their Wigner quasi-probability distributions is of significant and increasing interest \cite{Lvovsky2020}. Wigner-negativity is an immediate indicator of a uniquely quantum-mechanical state of light, and Wigner-negative states are a key resource in quantum computing \cite{Mari2012,Veitch2012,Veitch2013}, as well as in quantum metrology \cite{DemkowiczDobrzanski2009,Grochowski2025,Maliakal2026}. Most current methods for the generation of Wigner-negative light are inherently probabilistic, requiring heralding or other conditional measurements \cite{Ourjoumtsev2006,Gerrits2010,Huang2015,Baune2017,Eaton2019,Takase2022,Takase2023}.
Recently, an arguably simpler method was proposed by Quijandr\'ia et al. \cite{strandberg_one, strandberg_two}. Their method identifies Wigner-negative temporal modes in the steady-state output field of a coherently driven two-level system, with
states of these modes existing largely in the subspace spanned by the vacuum and single-photon Fock states. 
Their scheme was subsequently verified experimentally through the use of a circuit QED system \cite{strandberg_three}. More recently, this idea of considering the output field of a driven two-level system was extended in the theoretical work of \cite{miriam_thesis, leonhardt2025} to driving with a source of squeezed light. This was shown to lead to still more exotic states of the temporal modes involving higher photon numbers and resembling squeezed Schr\"odinger cat states.

In this article, we expand upon the methods used in these previous works by moving from the continuous-driving, steady-state regime to one of pulsed driving of a two-level system. We demonstrate, through numerical simulation, that temporal modes in the output field of a two-level system driven by a pulse of coherent or squeezed light can exhibit states similar to those from continuous driving, but with significantly increased levels of Wigner-negativity, and more pronounced non-classical distributions.

For the time dependence of the driving pulse and the output-field temporal mode, we choose Gaussians, and we investigate how the parameters of these Gaussians (i.e., their widths and relative offset) influence the state of the output-field temporal mode. 
Moreover, we use optimization techniques to determine the parameters that yield the maximum Wigner-negative volume. We also examine the similarities between the states generated with our method and other known quantum states, with particular focus on comparisons with displaced single-photon states for pulsed coherent driving, and squeezed Schr{\"o}dinger-cat-type states for pulsed squeezed driving. 

In addition to this, along a completely novel line of enquiry, we go beyond two-level systems to consider the pulsed driving of higher-dimensional spin systems. In particular, we study temporal modes present in the output field of pulse-driven spin-$S$ systems for $S\in\{ 1,3/2,2\}$. For coherent-state pulses we can make a connection with previous work on Wigner-negative states in the cavity output field of a continuously-driven Tavis-Cummings model of the same spin \cite{alex_negative}. 
However, for squeezed-state pulses incident upon a spin-1 system, we obtain an entirely new state of light in the outgoing temporal mode, with an intriguing structure of nonclassicality in its Wigner distribution.  
To offer some insight into the generation of this state, we also briefly examine the case of a spin-1 system driven with $N$-photon (i.e., Fock state) pulses, examining in particular the first-order (amplitude) correlation functions of the total output field in comparison with the temporal mode profile and noting a specific and tell-tale dependence of the correlation function on $N$.

A summary of the contents of the paper is as follows. 
In Section~\ref{section:model}, we introduce the physical configuration we are investigating and give the cascaded-systems master-equation that we use to describe the system. We also outline the methods used to select temporal modes of the driving and output fields  

In Section~\ref{section:coherent}, we consider the output field of a two-level (spin-1/2) system driven by a pulse of coherent light. We demonstrate the presence of strongly Wigner-negative temporal modes in the output field and point out advantages over the continuous driving regime. We examine the effect of changing the coherent field amplitude of the driving pulse on the selected temporal modes and optimize our system for the generation of the state of the mode with the highest Wigner-negative volume. Remarkably, we find that this state is very close to a displaced single-photon state. Then, we extend our model and optimization to emitters of higher spin, $S>1/2$, and obtain states with distinct similarities to displaced $(2S)$-photon states.

In Section~\ref{section:results}, we move to considering a source cavity containing a mode prepared in a squeezed state instead of a coherent state. After again noting the relative ease of generating strongly Wigner-negative states, we discuss what effect varying either the temporal mode parameters or the initial squeezing strength has on these modes, and optimize each parameter in order to find a maximally negative state. We also explore similarities with other quantum states.

In Section~\ref{section:spinone}, we investigate the entirely new dynamics involved in the pulsed squeezed driving of a spin-one system. We provide an initial look at the temporal modes of its output field, demonstrate its limitations in terms of generating Wigner-negativity, and describe its novel effect of excluding the two-photon component from the output field.

In Section \ref{section:correlation}, we discuss and investigate this two-photon exclusion within the context of the first-order correlation function of the outgoing field, and provide a framework for its action by analyzing the correlation functions of low-spin systems driven by $N$-photon pulses.
Finally, in Section \ref{section:conclusion}, we summarize our results.

\begin{figure}[htb]
    \centering
    \includegraphics[width=\linewidth]{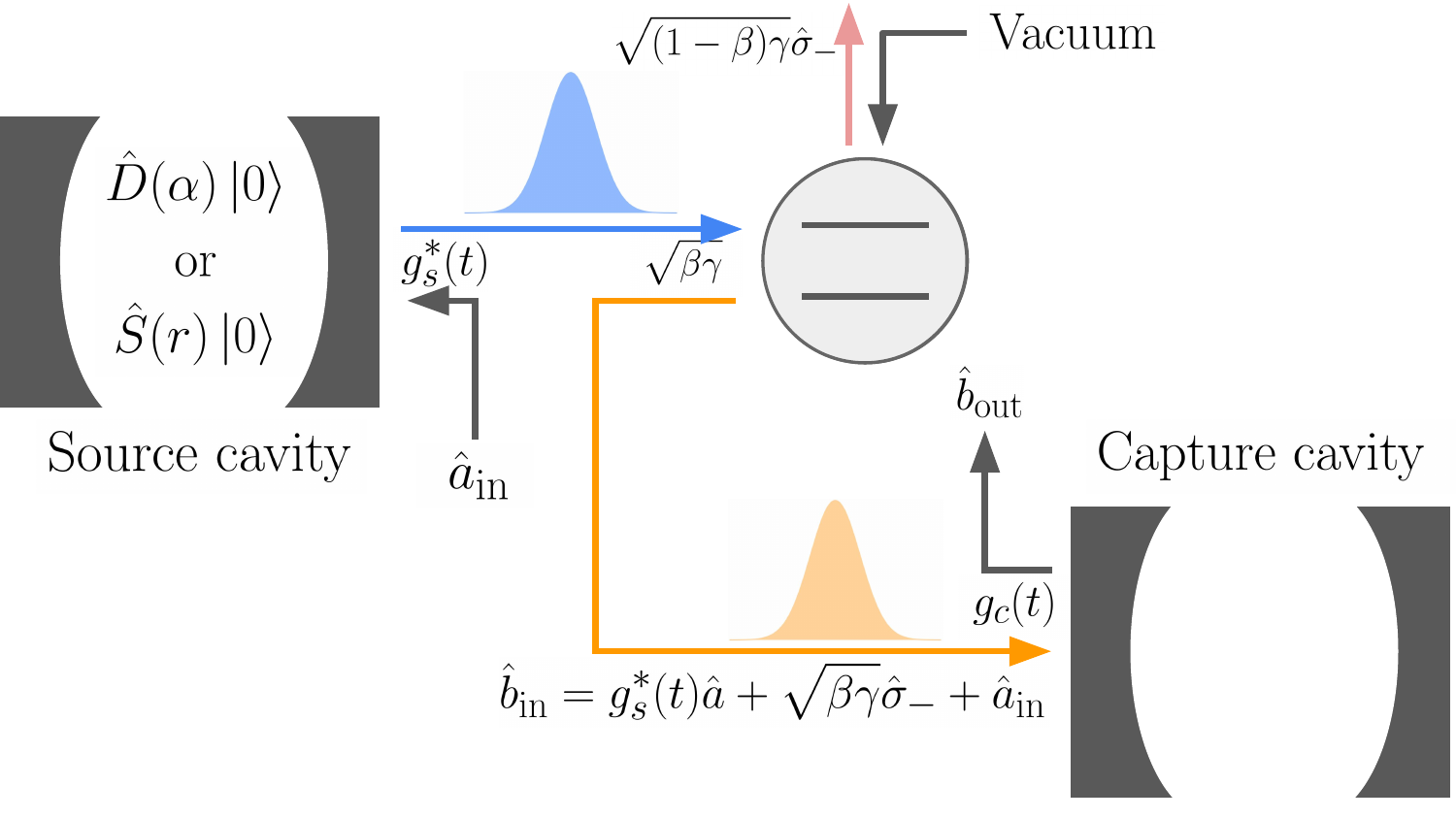}
    \caption{Diagram of the model we are simulating. A pulse of light is emitted from a source cavity, and cascaded through (in this instance) a two-level emitter into a capture cavity. We only consider a vacuum input $\hat{a}_\text{in}$ for the source cavity.}
    \label{figure:model_diagram}
\end{figure}

\section{Model}\label{section:model}

\subsection{The Cascaded Systems Master Equation}
The physical scenario we investigate is the output field of a two-level emitter driven by a pulse of light. 

In our model, depicted schematically in Fig.~\ref{figure:model_diagram}, we drive a two-level system with the output field of a cavity mode, which is initially assumed to be in either the coherent state
\begin{equation}\label{equation:definition_coherent}
    \ket{\alpha} = \hat{D}(\alpha)\ket{0} = \exp\left(\alpha \hat{a}^\dagger - \alpha^* \hat{a}\right)\ket{0},
\end{equation}
or the squeezed state
\begin{equation}\label{equation:definition_squeezed}
    \hat{S}(r)\ket{0} = \exp\left(\frac{r}{2}\left[\hat{a}^{\dagger^2} - \hat{a}^2\right]\right)\ket{0},
\end{equation}
where $\hat{a}$ is the annihilation operator of the quantized cavity mode.
The combined output field from the cavity and two-level system is then cascaded onto a second ``capture'' cavity, with a time-dependent coupling coefficient that is chosen to capture a specific temporal mode of the incident field. 

Using the cascaded-systems formalism \cite{cascaded_gardiner, cascaded_carmichael}, we fully describe the evolution of the system using the master equation
\begin{multline} \label{equation:definition_cascaded_master}
    \frac{d}{dt}\hat{\rho} = \frac{1}{i\hbar}\left[\hat{H}_S,\hat{\rho}\right] + \frac{1}{2}\mathcal{D}\left[\sqrt{\beta\gamma}\hat{\sigma}_- + g_{\text{s}}^*(t)\hat{a} + g_{\text{c}}^*(t)\hat{b}\right]\hat{\rho}  \\ + \frac{1}{2}\mathcal{D}\left[\sqrt{(1 - \beta)\gamma}\hat{\sigma}_-\right]\hat{\rho},
\end{multline}
 where, in a frame rotating with the cavity mode frequency, the system Hamiltonian is given by 
 \begin{multline} \label{equation:definition_cascaded_hamiltonian}
    \hat{H}_{\rm S} = \hbar \Delta \hat{\sigma}_+\hat{\sigma}_- + \frac{i\hbar}{2}\sqrt{\beta\gamma}\left(g_{\text{s}}(t) \hat{a}^\dagger \hat{\sigma}_- - g_{\text{s}}^*(t)\hat{a}\hat{\sigma}_+\right) \\ + \frac{i\hbar}{2}\left(g_{\text{s}}(t)g_{\text{c}}^*(t) \hat{a}^\dagger \hat{b} - g_{\text{s}}^*(t)g_{\text{c}}(t)\hat{a}\hat{b}^\dagger\right)\\ + \frac{i\hbar}{2}\sqrt{\beta\gamma}\left(g_{\text{c}}^*(t)\hat{\sigma}_+ \hat{b} - g_{\text{c}}(t)\hat{\sigma}_-\hat{b}^\dagger\right),
\end{multline}
with $g_{\text{s}}(t)$ being the time-dependent output coupling of the cavity mode [related to the line-width of the cavity by $g_{\text{s}}(t) = \sqrt{2\kappa(t)}\,$], $\hat{\sigma}_-$ the lowering operator of the two-level system, $\hat{b}$ the annihilation operator of the capture cavity mode, $g_{\text{c}}(t)$ the time-dependent coupling of the capture cavity mode, $\Delta$ the detuning of the two-level system from the cavity frequency, $\gamma$ the total decay rate of the two-level emitter, and $\beta$ the fraction of this decay that is coupled to the combined output field from the source cavity and the emitter, which is subsequently incident upon the capture cavity. 
The total input field to the capture cavity is thus described by the time-dependent field (annihilation) operator
\begin{equation} \label{equation:definition_output_field}
    \hat b_{\rm in}(t) = \sqrt{\beta\gamma}\hat{\sigma}_- + g_{\text{s}}^*(t)\hat{a} + \hat{a}_\text{in}(t),
\end{equation}
where $\hat{a}_\text{in}(t)$ represents the vacuum input field to the source cavity. 

In our simulations, we normalize all of our parameters to the decay-rate of the two-level emitter by setting $\gamma = 1$, which also scales the time used in our numerical evolution. Unless otherwise stated, we also set $\Delta = 0$ and $\beta = 1$, meaning the two-level emitter is driven on resonance and its entire output field is incident on the capture cavity. The effect of changing these assumptions is discussed later.

\subsection{Temporal Modes}

We are free in our model to have light emitted from the source cavity in any form of temporal mode through a suitable choice of $g_{\rm s}(t)$. However, for general applicability, and later largely justified by our results, we choose a pulse of light in a Gaussian temporal mode, defined by the (normalized) mode function 
\begin{equation} \label{equation:definition_gaussian}
    f_{\rm s}(t) = \left(\frac{8}{\pi \tau_{\rm s}^2}\right)^{1/4}\exp\left[-\left(\frac{t - t_{\rm s0}}{\tau_{\rm s}/2}\right)^2\right],
\end{equation}
with (temporal) full width at half maximum (FWHM) given by $T_\text{s} = \tau_{\rm s}\sqrt{\ln\left(2\right)}$, and where $t_{\rm s0}$ corresponds to the (temporal) centre of the mode. 
This temporal mode is produced by choosing the source-cavity output coupling to be
\begin{equation} \label{equation:definition_source_coupling}
    g_{\text{s}}(t) = \frac{f_{\rm s}^*(t)}{\sqrt{1 - \int_{-\infty}^t dt'\,\lvert f_{\rm s}(t')\rvert^2}} .
\end{equation}
Given that $\int_{-\infty}^\infty dt \lvert f_{\rm s}(t)\rvert^2 = 1$, 
the temporal mode operator defined as
\begin{equation} \label{equation:definition_temporal_mode}
    \hat{A}_{f_{\rm s}} = \int_{-\infty}^\infty dt\, f_{\rm s}(t)\hat{a}_{\rm out}(t) 
\end{equation}
then satisfies the standard bosonic commutation relation $[\hat A_{f_{\rm s}},\hat A_{f_{\rm s}}^\dag ]=1$ \cite{input_output_one, input_output_two}.

We investigate the propagating output field from the source-plus-emitter system using the so-called ``capture cavity'' method \cite{input_output_one, input_output_two}. This method makes use of the cascaded quantum systems formalism to introduce a virtual capture cavity at the output of our cascade. 
By choosing the input-output coupling strength of the capture cavity to have the time-dependent form 
\begin{equation} \label{equation:definition_capture_coupling}
    g_{\text{c}}(t) = \frac{-f_{\rm c}^*(t)}{\sqrt{\int_{-\infty}^t dt'\, \lvert f_{\rm c}(t')\rvert^2}},
\end{equation}
where $f_{\rm c}(t\rightarrow\infty )=0$ [and hence $g_{\rm c}(t\rightarrow\infty )=0$], 
the quantum state of the capture cavity mode at the end of the time evolution of the master equation is precisely that of the temporal mode defined by the mode function $f_c(t)$. For simplicity, we also choose this to be Gaussian,
\begin{equation} 
    f_{\rm c}(t) = \left(\frac{8}{\pi \tau_{\rm c}^2}\right)^{1/4}\exp\left[-\left(\frac{t - t_{\rm c0}}{\tau_{\rm c}/2}\right)^2\right] .
\end{equation}
As we are free to pick the start time of our simulation, we choose our source cavity temporal mode to be centered at $t=0$, i.e., $t_{\rm s0} =0$. We then have just three mode parameters to consider, $T_\text{s}$,  $T_\text{c}$, and $t_\text{c0}$.

To conclude this subsection, we emphasize that what we have just described is a purely theoretical method for determining the quantum state of a propagating temporal mode in the source-plus-emitter output field, but this state can in practice be measured experimentally using suitable (pulsed, mode-matched) homodyne detection and tomography. 
We note also that, in practice, the specific temporal modes that we highlight in this work typically contain only a portion of the total incident photon flux, but could be isolated (or separated) from the rest of the scattered output field using the quantum-pulse-gate (QPG) method \cite{Brecht2014,Serino2023}.

\begin{figure}[tb]
    \centering
    \includegraphics[width=\linewidth]{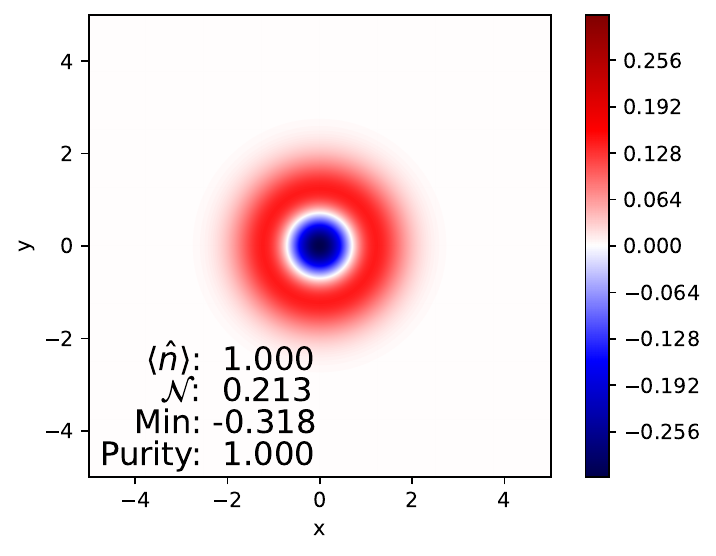}
    \caption{Wigner distribution of the single photon Fock state $\ket{1}$.}
    \label{figure:single_photon_wigner}
\end{figure}

\subsection{Wigner-Negative Volume}
\begin{figure*}[ht]
    \centering
    \includegraphics[width=\linewidth]{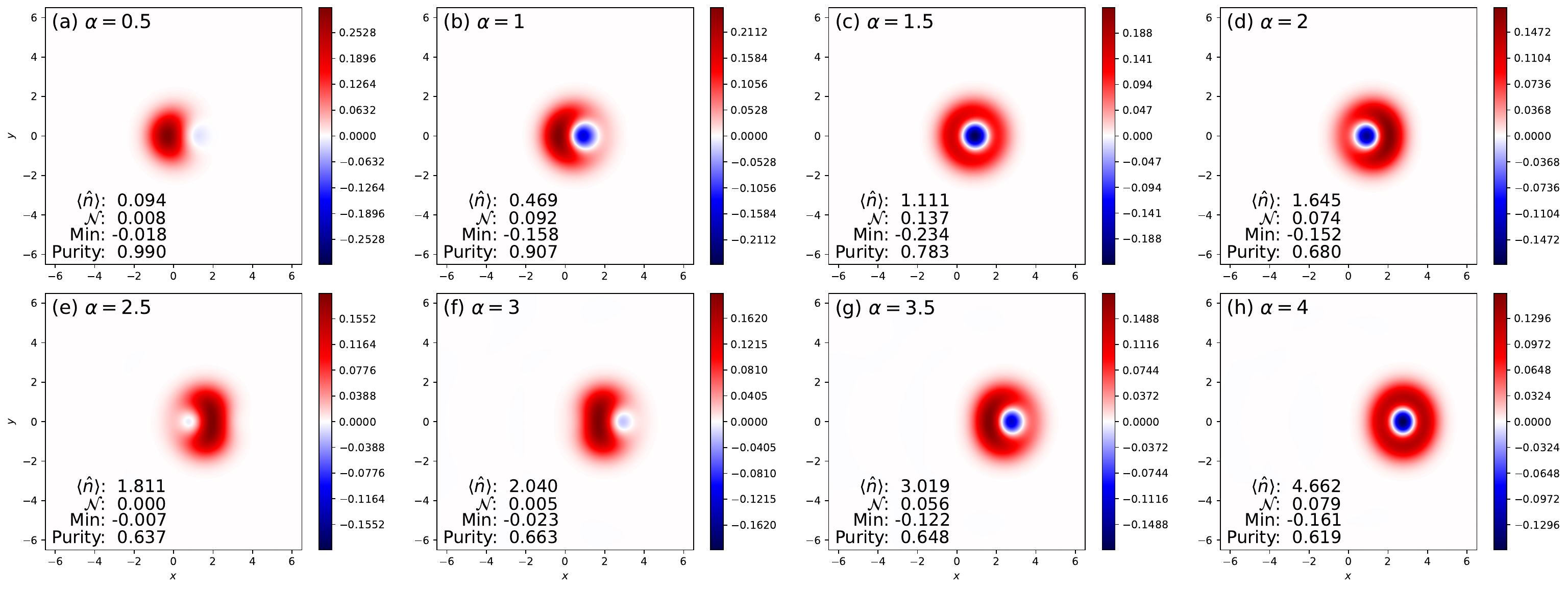}
    \caption{Wigner distributions for Gaussian temporal modes of the output field of a two-level system driven by a pulse of coherent light, for various initial coherent state amplitudes $\alpha$. Each simulation uses Gaussian temporal modes defined by 
    $(T_{\rm s},T_{\rm c},t_{\rm c0})=(1.11,2.55,1.15)\gamma^{-1}$. Note that the scale of the colour bar varies slightly between plots.}
    \label{figure:spin_half_coherent_variable_alpha_grid}
\end{figure*}
When considering a quantum state, there are two metrics with which we can quantify its Wigner-negativity. The first is the minimum value of the Wigner distribution, but this can be somewhat misleading if considering distributions with large regions of low but meaningful negativity. Hence, we instead focus mainly on the Wigner-negative volume \cite{negative_volume}, defined by
\begin{equation}\label{equation:definition_negative_volume}
    \mathcal{N} = \frac{1}{2}\iint dxdy\,\left\{ \lvert W(x,y)\rvert - W(x,y)\right\} ,
\end{equation}
as a quantitative measure of the degree of negativity of a quantum state.

To provide a benchmark for comparison, the Wigner distribution of the single photon Fock state, $\ket{1}$, given by 
\begin{equation}\label{equation:single_photon_wigner_distribution}
    W(x, y) = \frac{1}{\pi}e ^{-\left(x^2 + y^2\right)}\left(2x^2 + 2y^2 - 1\right),
\end{equation}
is shown in Fig.~\ref{figure:single_photon_wigner}.
This distribution has a negative volume $\mathcal{N} = 2e^{-1/2}-1=0.2131$ and a minimum value of $-0.3183$. 

Finally, we will also quote the purity, given by ${\rm Tr}(\hat\rho^2)$, for each state we present in our results. We note in advance that a general trend that we observe in our results is a decrease in purity of the state of the captured temporal mode with increasing strength (i.e., mean photon number) of the incident pulse. This is not unexpected, as increased intensity of the driving pulse leads to increased excitation of the emitter and, hence, a larger fraction of incoherent emission from the emitter in the total outgoing field.

\section{Pulsed coherent driving of a spin-$S$ emitter} \label{section:coherent}

\subsection{$S=1/2$}

Fig.~\ref{figure:spin_half_coherent_variable_alpha_grid} displays the Wigner distributions of output-field temporal modes of a two-level emitter driven by a pulse of coherent light. The driving pulse is a Gaussian with $T_\text{s} = 1.11\gamma^{-1}$, and the outgoing temporal mode considered is also Gaussian with $T_\text{c} = 2.55\gamma^{-1}$ and $t_\text{c0} = 1.15\gamma^{-1}$. The coherent field amplitudes of the driving pulse  vary in the range $ 0.5 \leq \alpha \leq 4$. These parameters will be justified when optimising for negativity later in this section. 

In most of the distributions, we observe pronounced Wigner-negativity, confirming the states as non-classical, and demonstrating a dramatic transformation from the incident Gaussian coherent state to highly non-Gaussian states in the output field.
The majority of these states show similarities to those found for the continuous-driving case considered previously \cite{strandberg_one,strandberg_two,strandberg_three}. That is, they consist of a single negative region partially surrounded by a positive ridge, which is characteristic of some admixture of the vacuum and single-photon states. 
However, this is not always the case as the coherent amplitude is varied here; for certain amplitudes, there is no visible negativity in the Wigner distribution ($\alpha = 2.5$), while for others the state is clearly much closer to a single-photon state ($\alpha = 1.5,3.5$), albeit displaced by some amount from the origin.

Also of note in Fig.~\ref{figure:spin_half_coherent_variable_alpha_grid} is that we observe states with negative volume $\mathcal{N}$ over 0.13, 
significantly larger (by a factor of more than $5$) than the optimum found with continuous driving in \cite{strandberg_one}.  
This is an important result in the context of the generation of propagating Wigner-negative field modes with a driven emitter, as it demonstrates that pulsed dynamics can substantially increase the overall Wigner-negativity associated with output-field temporal modes. In Section IV, we will find this to also be the case with squeezed driving of an emitter.

Let us now discuss further characteristics of the Wigner distributions shown in Fig.~\ref{figure:spin_half_coherent_variable_alpha_grid}. 
As the coherent amplitude $\alpha$ (taken here to be real) increases, the Wigner distribution as a whole moves to the right along the $x$-axis, as one might expect. 
However, the center location of the Wigner-negative region of the distribution (when it is present) does not vary in the same smooth fashion. Instead, it remains at essentially the same position ($x\simeq 1$) from $\alpha=0.5$ to $\alpha=2.0$, before, in the range $\alpha\simeq 2.5$ to $\alpha\simeq 3.0$, it first disappears and then reappears at a shifted center position ($x\simeq 3$). It follows obviously that the Wigner-negative volume does not vary monotonically with the driving amplitude. 

\begin{figure}
    \centering
    \includegraphics[width=\linewidth]{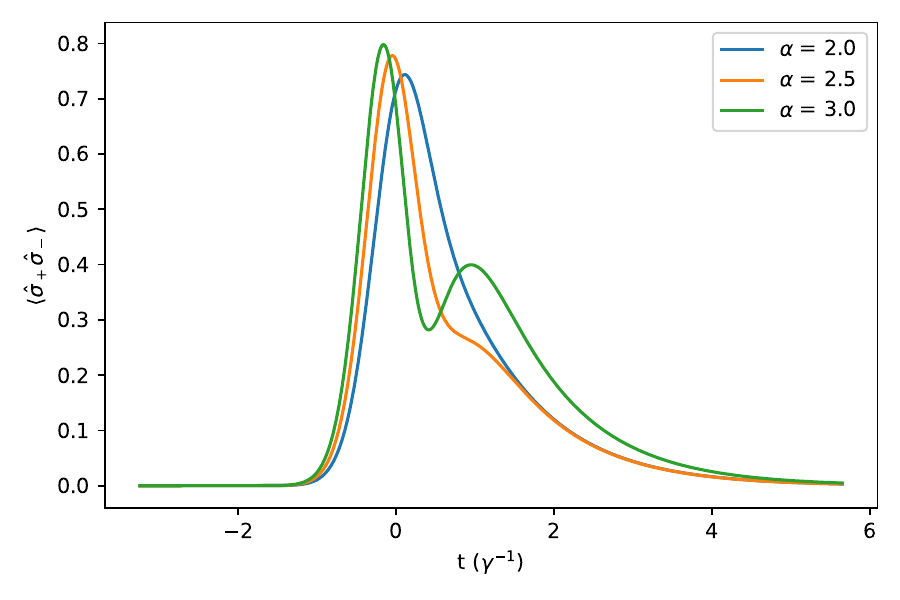}
    \caption{Expectation value $\langle \hat{\sigma}_+\hat{\sigma}_-\rangle$ for a spin-half emitter driven by a pulse of coherent light with a Gaussian temporal envelope of FWHM $T_{\rm c}=2.65\gamma^{-1}$, and coherent amplitude $\alpha \in \{2.0, 2.5, 3.0\}$.}
    \label{figure:spin_half_occupations}
\end{figure}

Some insight can be gained into these effects by considering the evolution of the excited state population of the emitter, quantified by the expectation value $\langle \hat{\sigma}_+\hat{\sigma}_-\rangle$. 
This population is shown as a function of time in Fig.~\ref{figure:spin_half_occupations} for the same incident driving pulse used in Fig.~\ref{figure:spin_half_coherent_variable_alpha_grid} and $\alpha \in \{2.0, 2.5, 3.0\}$.
The onset of Rabi oscillations in this range of $\alpha$ is evident and, 
in fact, we find that the first appearance of the Wigner-negative region at $x\simeq 3$ coincides with the first appearance of a second peak in the time variation of $\langle \hat{\sigma}_+\hat{\sigma}_-\rangle$.

We also draw attention to the mean photon number of each shown output field mode, and namely that in each case it is only a small fraction of the mean photon number of the input coherent state, ranging from 49.4\% when $\alpha=1.5$ to only 22.7\% when $\alpha = 3$. Motivated by this, we have also investigated the states of other orthogonal temporal modes to the Gaussian modes shown in Figure \ref{figure:spin_half_coherent_variable_alpha_grid}, however, despite containing relatively large photon number fractions, other orthogonal modes display zero Wigner negativity, and appear to contain approximately coherent states.

Although a complete exploration of the parameter space is not practical due to the numerical complexity of solving the master equation, maximizing the negative volume through optimization of parameters remains possible. Using a variant of the Fuzzy Self-Tuning Particle Swarm Optimisation (FSTPSO) algorithm described in \cite{FSTPSO}, we believe we have found the state of maximum negative volume for this particular system.

\begin{figure}[ht]
    \centering
    \includegraphics[width=0.85\linewidth]{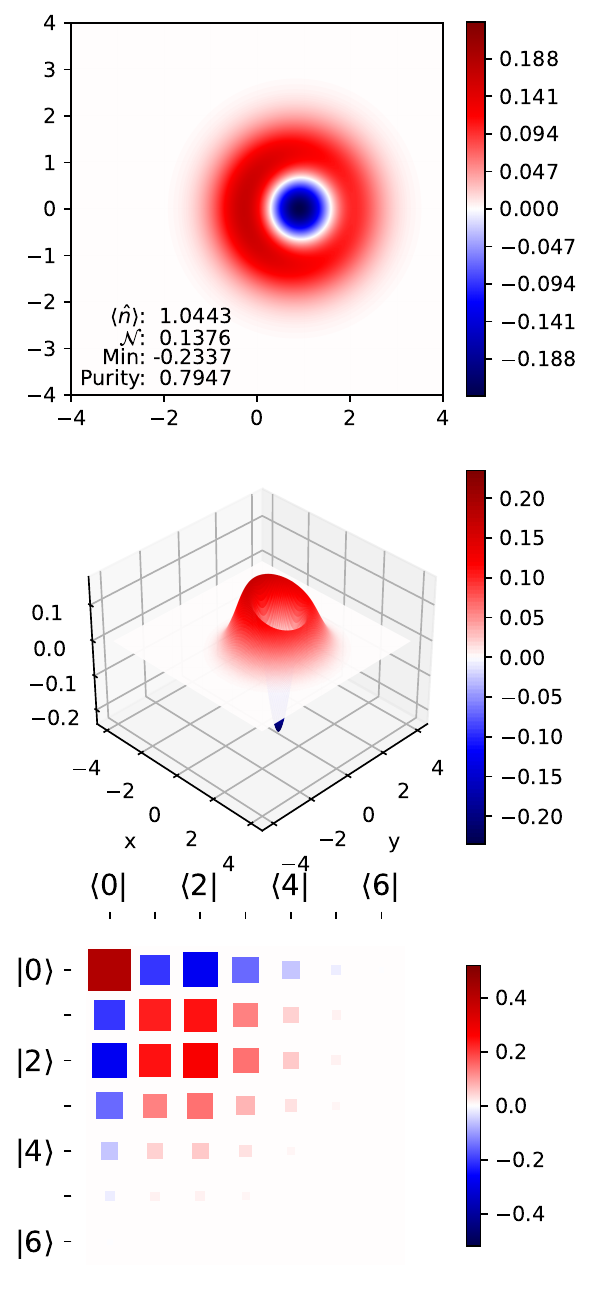}
    \caption{Wigner distribution (top and centre) and Hinton plot (bottom) for the captured state of maximum Wigner-negative volume, using $\alpha = 1.4503$, 
    $(T_{\rm s},T_{\rm c},t_{\rm c0})=(1.1146,2.5493,1.1502)\gamma^{-1}$. The Hinton plot has been truncated to the 6 photon state for visibility.}
    \label{figure:spin_half_coherent_max_N_triple}
\end{figure}

The Wigner distribution of this state, along with a Hinton plot of the corresponding density operator (matrix), is shown in Fig.~\ref{figure:spin_half_coherent_max_N_triple}. It possesses a Wigner-negative volume of $\mathcal{N} = 0.1376$, using an incident coherent-state pulse with $\alpha = 1.37$, and Gaussian temporal modes defined by $(T_{\rm s},T_{\rm c},t_{\rm c0})=(1.1146,2.5493,1.1502)\gamma^{-1}$.
An important caveat to acknowledge is that we have only considered Gaussian temporal modes in our optimization. In fact, optimization with more general temporal modes reveals that somewhat higher negative volumes can be achieved. However, limited experimentation with other temporal modes defined by analytical functions has not found any improvement.

\begin{figure*}[ht]
    \centering
    \includegraphics[width = \linewidth]{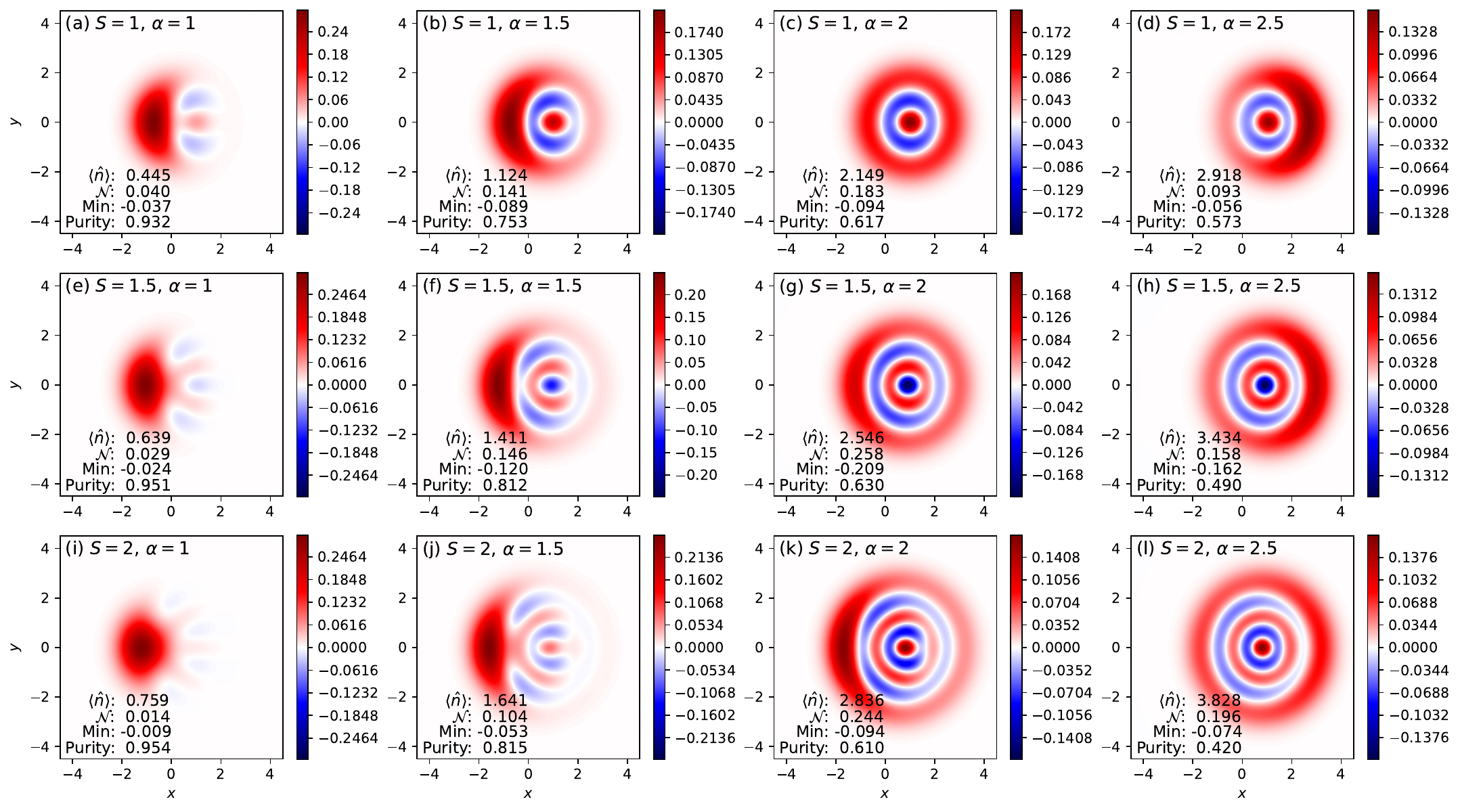}
    \caption{Wigner distributions for Gaussian temporal modes in the output field of an emitter of spin $S \in \{1, 1.5, 2\}$ driven by a pulse of coherent light of various initial coherent state amplitude $\alpha \in \{1, 1.5, 2, 2.5\}$. Each simulation uses Gaussian temporal modes defined by
    $(T_{\rm s},T_{\rm c},t_{\rm c0})=(0.7,1.35,0.7)\gamma^{-1}$. Note that the scale of the color bar varies slightly between plots.}
    \label{figure:spin_variable_coherent_variable_alpha_grid}
\end{figure*}
Finally, following on from our earlier comment, the optimized state shown in Fig.~\ref{figure:spin_half_coherent_max_N_triple} is clearly very close to a displaced single-photon state. Removing the finite coherent amplitude ($\langle \hat{a}\rangle = 0.4469$) from the state via application of the coherent displacement operator, we obtain a state that has a fidelity of over 0.96 with a pure single-photon state. 
Furthermore, preliminary testing has shown that if we allow for the use of completely general, non-Gaussian temporal modes, we can achieve fidelities of over 0.99.
This makes our system very promising as a deterministic source of  displaced single-photon states, from which pure single-photon states could also be extracted.

\subsection{$S>1/2$}
Recently, sources of Wigner-negative light involving the coherent driving of systems of higher total spin have been considered, for example in the form of Tavis-Cummings model, as in \cite{alex_negative}. This becomes significantly harder to numerically simulate, as the dimensionality of the system rapidly increases. However, results can still be obtained for some range of spin lengths. Based on the previous section, and the efficacy of pulsed coherent driving of a two-level, i.e., spin-1/2, emitter, we now extend our methodology to consider the pulsed coherent driving of a spin-$S$ emitter with $S>1/2$. 
Within our model, this simply amounts to replacing the $\hat{\sigma}_+$ and $\hat{\sigma}_-$ operators with general spin raising and lowering operators $\hat{S}_+$ and $\hat{S}_-$, which act on states of higher-dimensional angular momentum. 

\begin{figure}
    \centering
    \includegraphics[width=\linewidth]{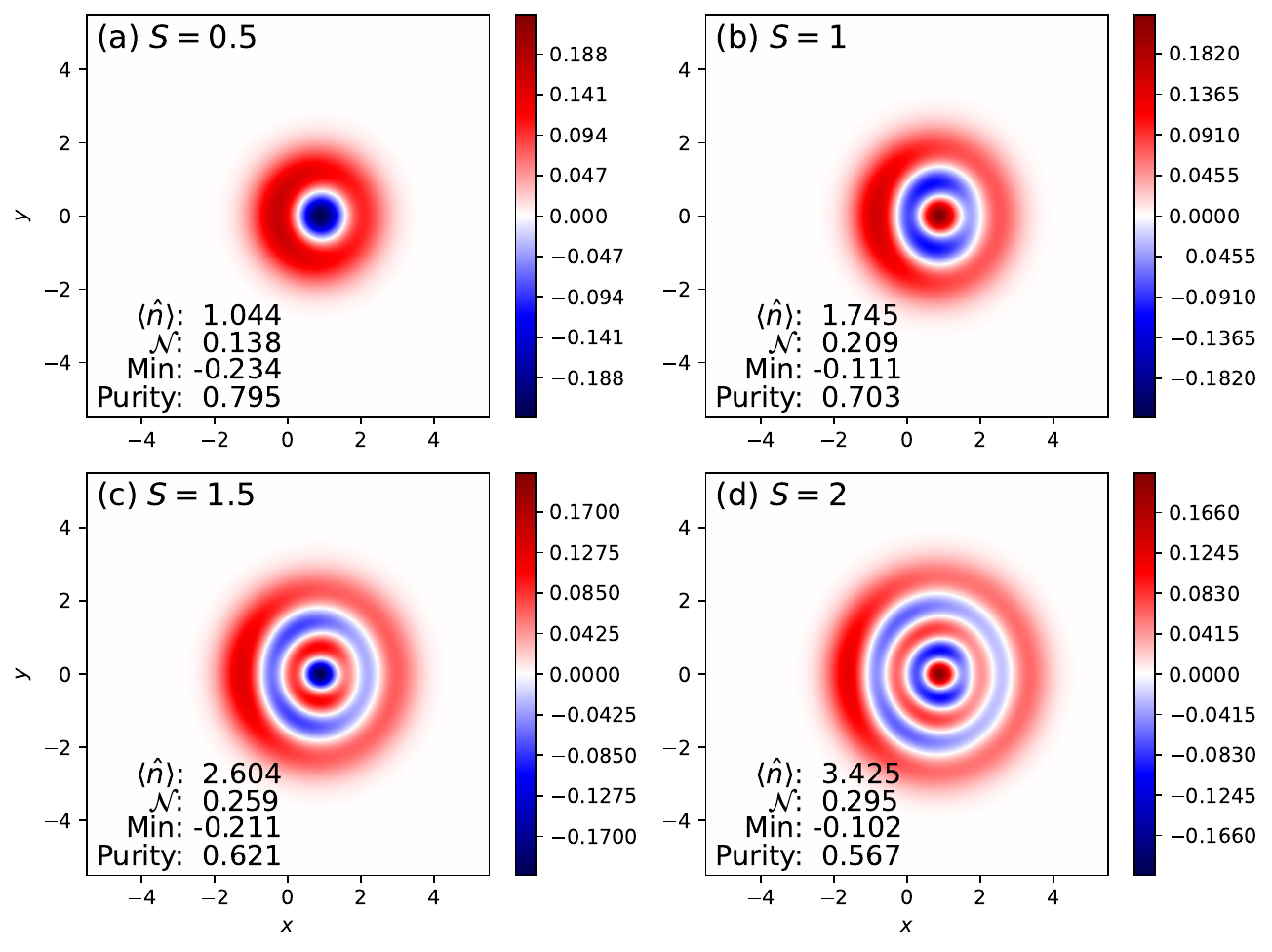}
    \caption{Wigner distributions for the captured states of maximum negative volume for driven emitters of different spin $S$. Each state is the maximum negative volume state obtained through optimization of the initial coherent amplitude and Gaussian temporal mode parameters. The parameters for each state are:\\ 
    (a) $\alpha = 1.4503$, $(T_{\rm s},T_{\rm c},t_{\rm c0})=(1.1146,2.5493,1.1502)\gamma^{-1}$,
    \\
    (b) $\alpha = 1.6752$, $(T_{\rm s},T_{\rm c},t_{\rm c0})=(1.0163,1.8027,0.8876)\gamma^{-1}$,
    \\ 
    (c) $\alpha = 2.0250$, $(T_{\rm s},T_{\rm c},t_{\rm c0})=(0.7062,1.3571,0.7130)\gamma^{-1}$,
    \\
    (d) $\alpha = 2.2731$, $(T_{\rm s},T_{\rm c},t_{\rm c0})=(0.5828,1.0808,0.5998)\gamma^{-1}$.}
    \label{fig:coherent_higher_spins}
\end{figure}

\begin{figure*}[t]
    \centering
    \includegraphics[width=\linewidth]{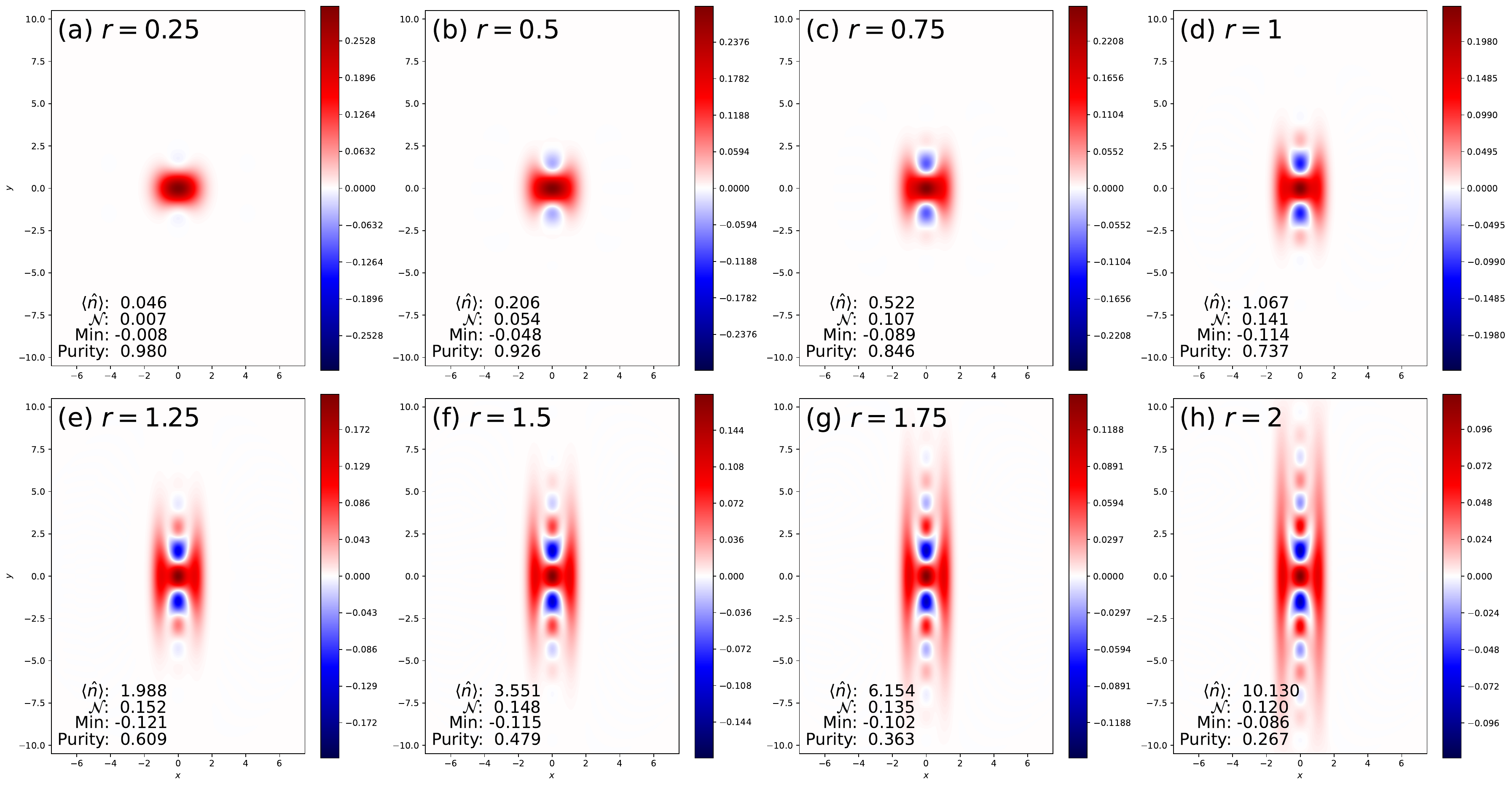}
    \caption{Wigner distributions for Gaussian temporal modes of the output field, for varied initial squeezed state squeezing parameters $r$. Each simulation uses Gaussian temporal modes defined by 
    $(T_{\rm s},T_{\rm c},t_{\rm c0})=(1.60,1.86,0.41)\gamma^{-1}$. Note that the scale of the colour bar varies slightly between plots.}
    \label{figure:spin_half_squeezed_variable_r_grid}
\end{figure*}
Many of the same patterns, with regards to structure and negative volume, exist for these higher-spin systems; specifically, the pattern of growing coherent amplitude and consequent ``shifting'' of the Wigner distribution along the $x$-quadrature, along with distributions of generally lower magnitude. As such, here we focus on the differences between states generated with different spins and the maximum negative volumes obtained for each spin, where again all optimization has been performed using the same FST-PSO variant algorithm. A selection of these states, from $S = 1/2$ to $S = 2$, along with the relevant coherent state and Gaussian temporal mode parameters, is included in Fig.~\ref{fig:coherent_higher_spins}. 

We immediately note the large negative volumes of the generated states, this time reaching over $\mathcal{N} = 0.25$ for the pulsed coherent driving of a spin-3/2 emitter. 
The observed increase in the maximum negative volume with increasing spin shows promise for the driving of emitters of even higher spin, and the volumes obtained far exceed those found in any other similar deterministic methods we have seen or considered.

Furthermore, the pattern of changing structure with increasing spin is extremely similar to that of pure $N$-photon states. 
Although further optimisation and investigation of this effect is outside the scope of this paper, this is extremely promising for the potential deterministic generation of displaced $N$-photon states, from which pure $N$-photon states could potentially be recovered by a suitable coherent displacement.

As expected based on the spin-1/2 case, these states show strong similarities to temporal modes present in the continuous coherent driving of low-spin systems, such as the Tavis-Cummings type model in \cite{alex_negative}. These models show the same pattern of an increasing number of negative lobes placed within a positive shell, but the Tavis-Cummings type states do not display nearly as clear similarities with displaced $N$-photon states as those we consider. Further testing has shown we can generate states more similar in structure to the Tavis-Cummings model with our system, but with significantly greater negativity, simply by adjusting the parameters of our simulations.

\section{Pulsed squeezed driving of a two-level emitter}\label{section:results}

\subsection{Wigner-negativity}
We now move to considering the pulsed \emph{squeezed} driving of a two-level emitter, motivated in this case by the states of temporal modes that were found when considering continuous squeezed driving \cite{miriam_thesis, leonhardt2025}. 
As mentioned earlier, and in keeping with the results of the previous section, we find similar states with pulsed driving, but with much-enhanced features and negativity in their Wigner distributions compared to the case of continuous driving.

Fig.~\ref{figure:spin_half_squeezed_variable_r_grid} shows Wigner distributions of output field modes, for Gaussian source and capture cavity modes defined by $(T_{\rm s},T_{\rm c},t_{\rm c0})=(1.65,1.87,0.425)\gamma^{-1}$, for variable initial squeezing parameters $r$. This particular choice of parameters will again be justified later through optimization. 
We see strongly Wigner-negative states displaying the same general structure as those in \cite{miriam_thesis, leonhardt2025}, but with much greater Wigner-negative volumes. In particular, these reach values over $\mathcal{N} = 0.15$, already nearly matching the greatest Wigner-negative volume we found earlier for pulsed coherent driving of a spin-1 emitter. 

At high initial squeezing parameters, we also see a clear extension of the pattern found in the continuous driving case, where increased squeezing of the driving field results in more pairs of positive and negative lobes appearing along the $y$-axis. Up to three pairs of negative lobes are visible in the Wigner distribution for $r > 1.5$. 

\begin{figure}[t]
    \centering
    \includegraphics[width=\linewidth]{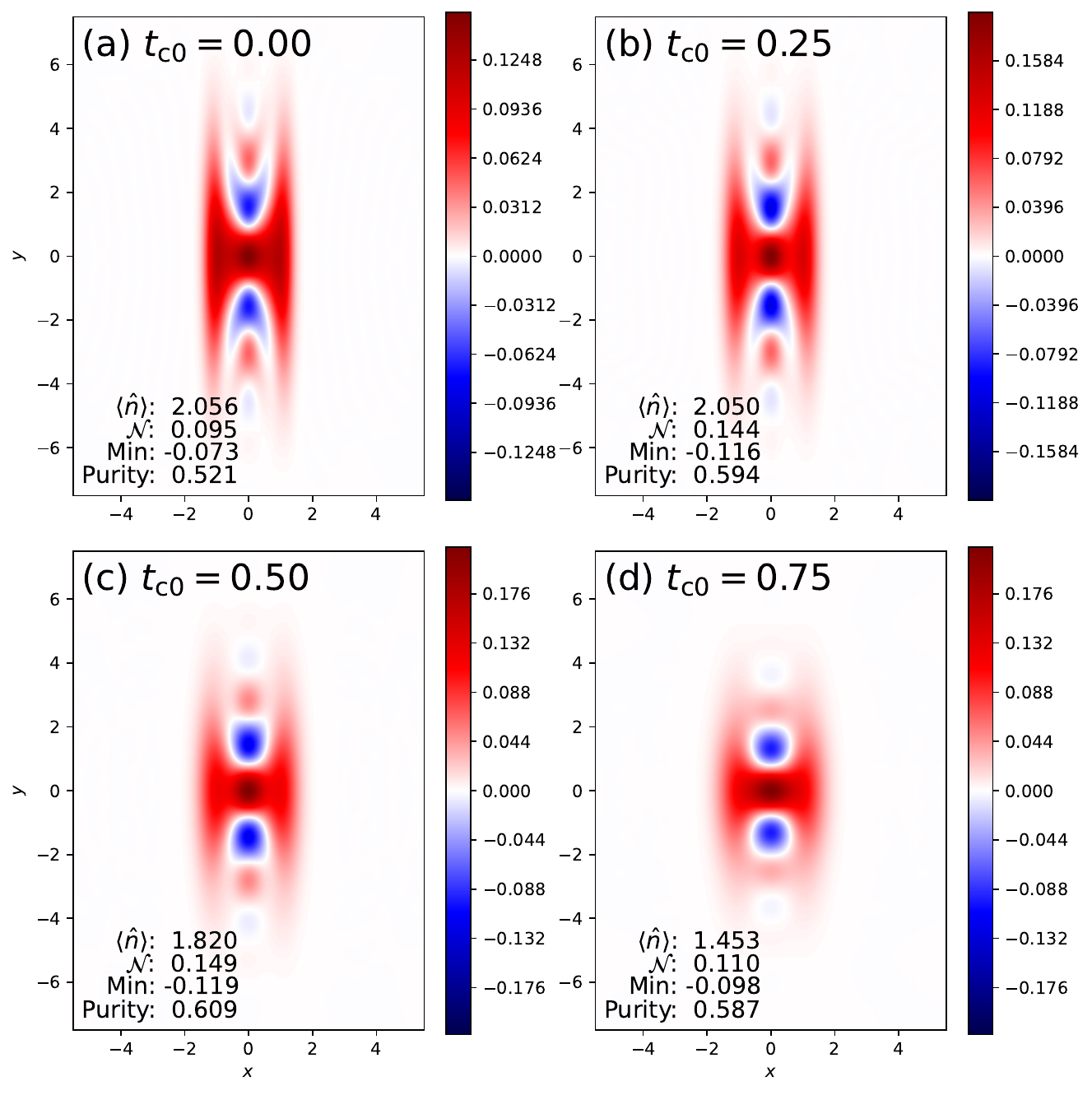}
    \caption{Wigner distributions for Gaussian temporal modes of the output field, for varied temporal offset $t_\text{c0}$ between the source and capture cavity temporal modes. Each simulation has used an initial squeezed state with $r=1.25$, and Gaussian temporal modes defined by $T_\text{s} = 1.60\gamma^{-1}$, $T_\text{c} = 1.86\gamma^{-1}$, and $t_\text{c0}\in \{0, 0.25\gamma^{-1},0.50\gamma^{-1},0.75\gamma^{-1}\}$. Note that the scale of the colour bar varies slightly between plots.}
    \label{figure:spin_half_squeezed_variable_tc0_grid}
\end{figure}

Otherwise, many of the same trends appear present for both pulsed and steady-state driving regimes. The density matrix describing the state of the output temporal mode lacks any contribution from coherences of the form $\bra{2n+1}\hat{\rho}\ket{2m}$ or $\bra{2n}\hat{\rho}\ket{2m+1}$, where $n, m\in\{0,1,2,\dots\}$. 
The most significant contributions come from the matrix elements $\bra{2n}\hat{\rho}\ket{2m}$, which one would largely expect due to the squeezed state input containing only even numbers of photons. However, faint contributions from elements $\bra{2n+1}\hat{\rho}\ket{2m+1}$ are now also present. 

We also see an initial trend of increasing negative volume with increasing $r$, before reaching a maximum near $r = 1.25$ and then decreasing for greater values of $r$. The minimum values of each distribution follow the same pattern, whilst the purity of the captured state decreases monotonically as $r$ increases. All of these trends are similar to those observed in the steady-state driving regime, although they vary from the pulsed coherent driving case.

Other than the appearance of additional negative lobes, further qualitative differences occur in the structure of each Wigner distribution as $r$ increases. The magnitude of the central maximum decreases continuously, and the two positive ridges stretch further along the $y$-axis. There are also subtle changes to the shape of the lobes, but these can also be achieved by changing the mode parameters, and thus it is unclear whether this is an effect of higher squeezing or the specific modes used. For example, the effect of varying $t_\text{c0}$ on the state depicted in Fig.~\ref{figure:spin_half_squeezed_variable_r_grid}(e) is shown in Fig.~\ref{figure:spin_half_squeezed_variable_tc0_grid}. As $t_\text{c0}$ increases, capturing a temporal mode further displaced in time from the source pulse, the ridges of the Wigner distribution become less squeezed, and the shape of each lobe becomes more rounded. 

The continued use of Gaussian modes for our results is justified through a spectral decomposition of the first order correlation function of the output field, as described in \cite{input_output_two}, and discussed later in Section~\ref{section:correlation}. This spectral analysis reveals strong overlap of the Gaussian temporal modes with the dominant mode function of the output field. 
We have, in addition, trialled other analytic options for both the source- and capture-cavity temporal modes, such as box or $g^{(1)}$ filters. However, while each of these trials produced qualitatively similar states, they yielded significantly lesser negative volumes compared to the Gaussian modes. As with the spin-1/2 case, higher negative volumes have been found with more general temporal modes, but no analytically-defined modes have been found displaying greater negative volumes.

\begin{figure}[ht]
    \centering
    \includegraphics[width=0.9\linewidth]{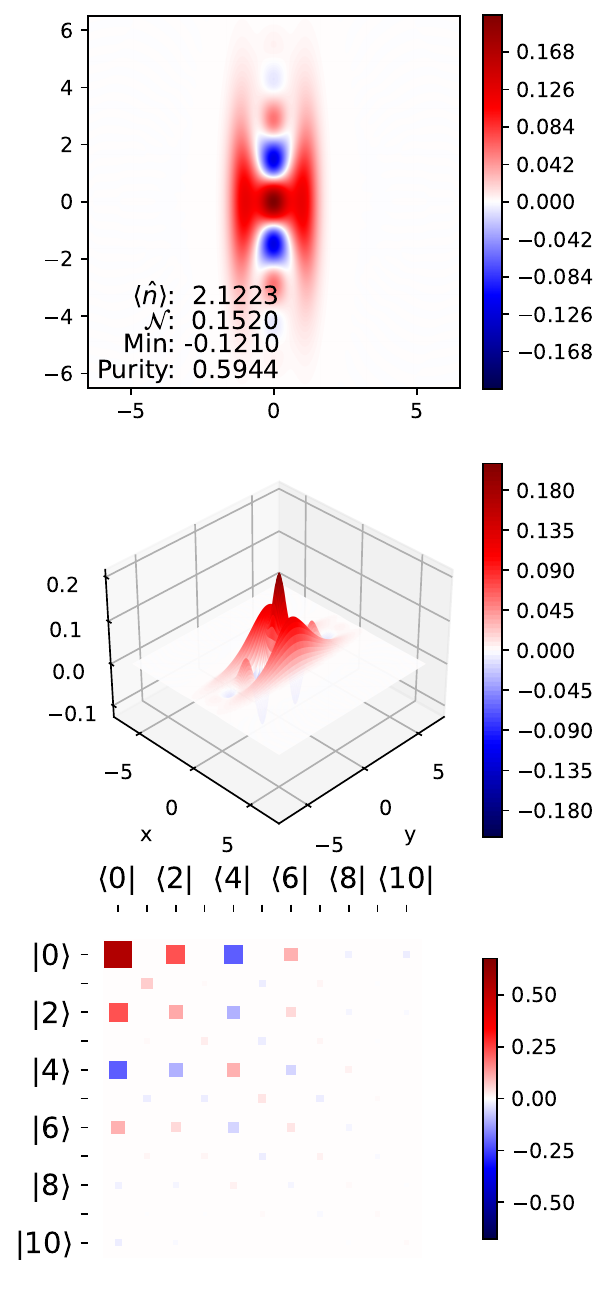}
    \caption{Wigner distribution (top and centre) and Hinton plot (bottom) for the temporal mode state of maximum Wigner negative volume, using $r = 1.2774$, $(T_{\rm s},T_{\rm c},t_{\rm c0})=(1.5973,1.8594,0.4087)\gamma^{-1}$. The Hinton plot has been truncated at the 10-photon state for visibility.}
    \label{figure:spin_half_squeezed_max_N_triple}
\end{figure}

We consider again an optimization of the initial squeezed state and Gaussian mode parameters to find the state of maximum possible negative volume. The Wigner distribution of this state, along with its Hinton plot, is shown in Fig.~\ref{figure:spin_half_squeezed_max_N_triple}. It has a Wigner-negative volume of $\mathcal{N} = 0.1520$, for an initial squeezed state with $r = 1.2609$, and Gaussian temporal modes defined by $(T_{\rm s},T_{\rm c},t_{\rm c0})=(1.6442,1.8715,0.4129)\gamma^{-1}$.
While this negative volume is larger than any generated from pulsed coherent driving of a spin-1/2 emitter, it does not exceed the maximum volume obtained for the pulsed coherent driving of a spin-1 system. 
However, the unique structure of the state in Fig.~\ref{figure:spin_half_squeezed_max_N_triple} makes it of considerable interest in its own right, as we will discuss shortly.

Finally, it is also important to note that these parameters are not optimal for observing Wigner-negative volumes for any input state, and the values of $T_\text{s}$, $T_\text{c}$, and $t_\text{c0}$ that result in the state of greatest Wigner-negative volume depend on the initial squeezing parameter of the input state. Generally, lesser-squeezed initial states require broader temporal modes than more highly squeezed initial states to maximize the negative volume.

\subsection{Effect of detuning and free-space emission}
In reality, free-space spontaneous emission and other effects will always result in some portion of the total output field not being incident on the capture cavity in our model (and thereby not contributing to the temporal mode of interest). We can consider a more physically realistic system by reintroducing an additional decay channel from the two-level emitter in our model. Computationally, this amounts simply to setting $\beta < 1$ in Eq.~(\ref{equation:definition_cascaded_master}).

Similarly, we can examine what happens if the two-level emitter is detuned from the source cavity mode by setting $\Delta \neq 0$. As either of these modifications increases the computational cost of running each simulation, we have focused on rigorously demonstrating the effect they have on the negative volume for a low squeezing parameter of the incident state, and assert the effect remains similar for larger squeezing.

\begin{figure}[h]
    \centering
    \includegraphics[width=0.9\linewidth]{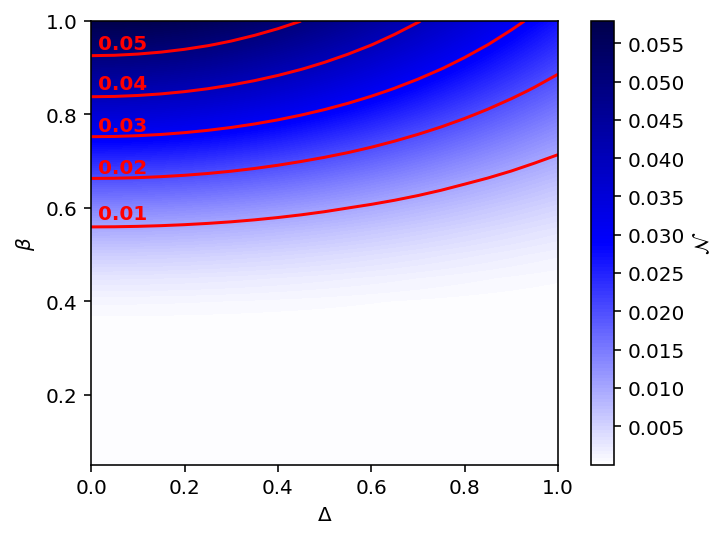}
    \caption{Contour plot displaying the effect of modifying the $\beta$ and $\Delta$ parameters on the observed negative volume $\mathcal{N}$. All simulations use an initial squeezed state with $r = 0.5$, and Gaussian temporal modes defined by $T_\text{s} = T_\text{c} = 2.15\gamma^{-1}$ and $t_\text{c0} = 0.6\gamma^{-1}$. Contour lines show negative volume increments of 0.01, from $\mathcal{N} = 0.01$ to $\mathcal{N} = 0.05$. The maximum negative volume is $\mathcal{N} = 0.054$, at $\Delta = 0$ and $\beta = 1$.}
    \label{figure:beta_delta_contour}
\end{figure}

We see in Fig.~\ref{figure:beta_delta_contour} that, not surprisingly, reducing $\beta$ from one or increasing $\Delta$ from zero decreases the captured negative volume. However, a reasonable fraction of the maximum negative volume can still be captured as long as $\beta\gtrsim 0.8$. We note that $\Delta$ and $\beta$ affect the state of the temporal mode in different ways; $\Delta>0$ introduces a partial rotation (i.e., component phase shift) of the Wigner distribution, while reducing $\beta$ results in a growing contribution of the vacuum state. 

\begin{figure*}[t]
    \centering
    \includegraphics[width=\linewidth]{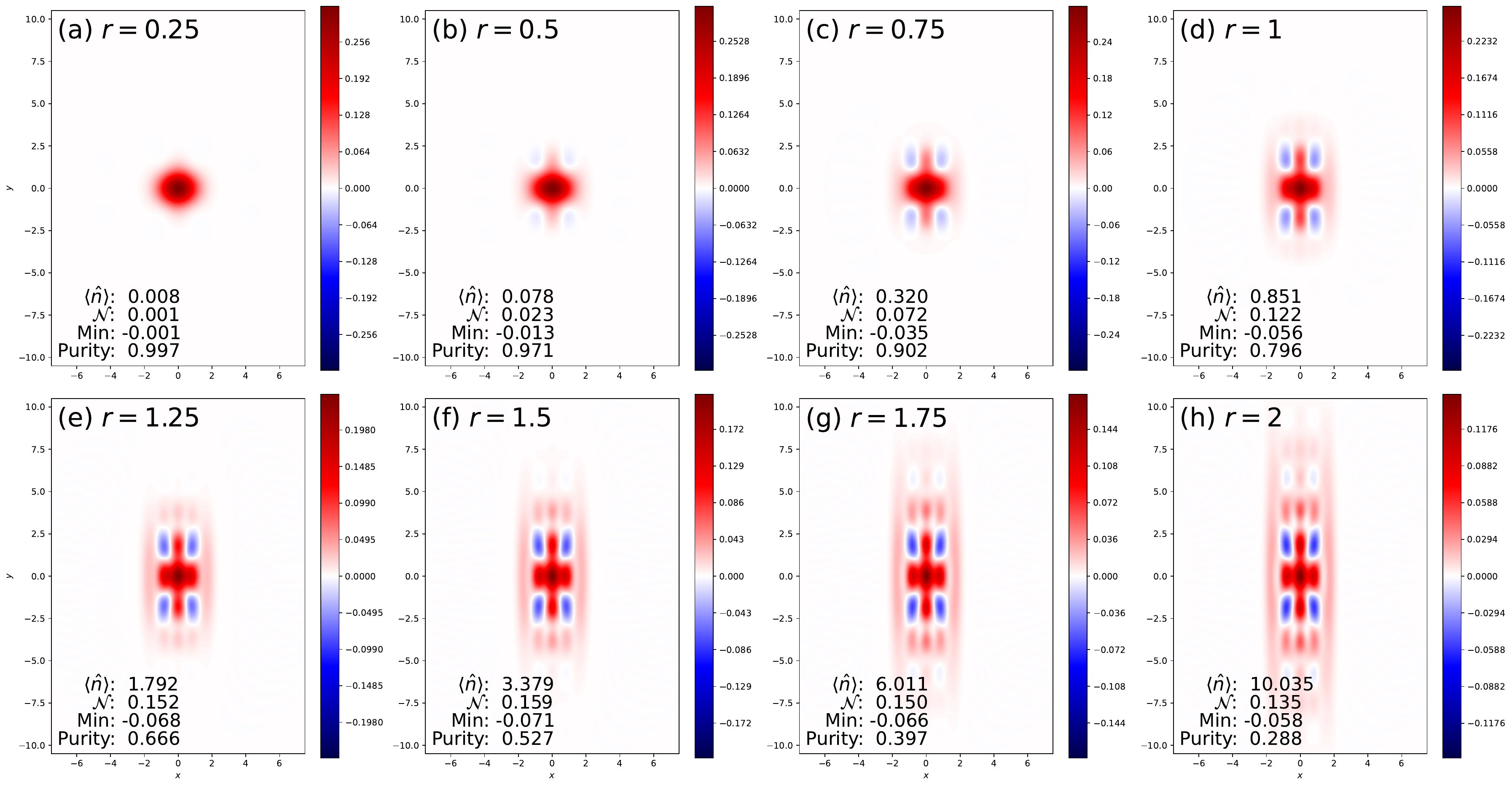}
    \caption{Wigner distributions for Gaussian temporal modes of the output field when driving a spin-1 emitter, for varied initial squeezing parameters $r$. The simulations use Gaussian temporal modes defined by $(T_{\rm s},T_{\rm c},t_{\rm c0})=(0.93, 0.97, 0.22)\gamma^{-1}$. Note that the scale of the colour bar varies slightly between plots.}
    \label{figure:spin_one_squeezed_variable_r_grid}
\end{figure*}

\subsection{Comparison with squeezed Schr{\"o}dinger Cat  states} 

As found in the steady-state driving regime, the temporal mode states on display in Figs.~\ref{figure:spin_half_squeezed_variable_r_grid}-\ref{figure:spin_half_squeezed_max_N_triple} show distinct similarities with squeezed Schr{\"o}dinger cat (SSC) states; i.e., with superpositions of displaced squeezed states of the form
\begin{equation}\label{equation:definition_schrodinger_cat}
    \ket{\psi}_{\rm SSC} = N \left(\hat{D}(\alpha') + \hat{D}(-\alpha')\right)\hat{S}(r')\ket{0},
\end{equation}
for some displacement and squeezing parameters $\alpha'$ and $r'$, respectively, with $N$ a normalisation factor. 
As in \cite{leonhardt2025}, we can use the fidelity \cite{Jozsa1994},
\begin{equation}\label{equation:definition_fidelity}
    F(\hat{\rho}_1,\hat{\rho}_2) = \left(\text{Tr}\left[ \sqrt{\sqrt{\hat{\rho}_1} \hat{\rho}_2\sqrt{\hat{\rho}_1}}\right]\right),
\end{equation}
to quantify the degree of similarity between the temporal mode states and the state (\ref{equation:definition_schrodinger_cat}). 

\begin{table}[h]
    \centering
    \begin{tabular}{c|ccc}
        $r$ & $\alpha'$ & $r'$ & $F$ \\
        \hline
        0.25 & 0.5097 & 0.1774 & 0.9952\\
        0.50 & 0.6176 & 0.3442 & 0.9816\\
        0.75 & 0.6631 & 0.4887 & 0.9589\\
        1.00 & 0.6873 & 0.6122 & 0.9242\\
        1.25 & 0.7036 & 0.7241 & 0.8764\\
        1.50 & 0.7170 & 0.8351 & 0.8180\\
        1.75 & 0.7289 & 0.9500 & 0.7538\\
        2.00 & 0.7385 & 1.0570 & 0.6884
    \end{tabular}
    \caption{Optimum fidelities between the states shown in Fig.~\ref{figure:spin_half_squeezed_variable_r_grid} and a SSC-state defined by Eq.~(\ref{equation:definition_schrodinger_cat}), along with the corresponding SSC-state parameters $\alpha'$ and $r'$ for each state.}
    \label{table:spin_half_squeezed_cat_comparison_table}
\end{table}

Table~\ref{table:spin_half_squeezed_cat_comparison_table} displays the optimum fidelities between the states shown in Fig.~\ref{figure:spin_half_squeezed_variable_r_grid} and a SSC-state. We find high fidelities for most of the range of squeezing parameters considered; above $0.9$ for $r\leq 1$, and still significant  for greater initial squeezing parameters. They are comparable to those found in the steady-state driving regime \cite{leonhardt2025}, especially near where the maximum negative volumes are obtained in each regime, and highlight the potential of our pulsed system as a deterministic source of propagating SSC-state modes of light. 
These are, of course, also of considerable interest in the context of the preparation of Gottesman-Kitaev-Preskill (GKP) states \cite{GKP}, and preliminary analysis involving evaluation of GKP stabilizers shows promising results.

\section{Pulsed squeezed driving of a spin-1 emitter} \label{section:spinone}

It is natural to consider the extension of our investigation to pulsed squeezed driving of an emitter of higher spin, e.g., a spin-1 emitter. 
Fig.~\ref{figure:spin_one_squeezed_variable_r_grid} shows Wigner distributions for output-field temporal modes with variable initial squeezing parameter and a spin-1 emitter. The simulations use Gaussian temporal modes defined by $(T_{\rm s},T_{\rm c},t_{\rm c0})=(0.93, 0.97, 0.22)\gamma^{-1}$. 

The distributions are dramatically modified from the SSC-type states obtained for a spin-1/2 emitter and possess a distinct, grid-like structure, somewhat reminiscent of finite-energy GKP states (for sufficiently large $r$). 
In terms of Wigner-negativity, the obtained distributions for the temporal mode states have minima that are notably less negative (by a factor of roughly one half) than the spin-1/2 case, but they possess comparable negative volumes, with a maximum $\mathcal{N} \simeq 0.16$. 

\begin{figure}[t]
    \centering
    \includegraphics[width=0.9\linewidth]{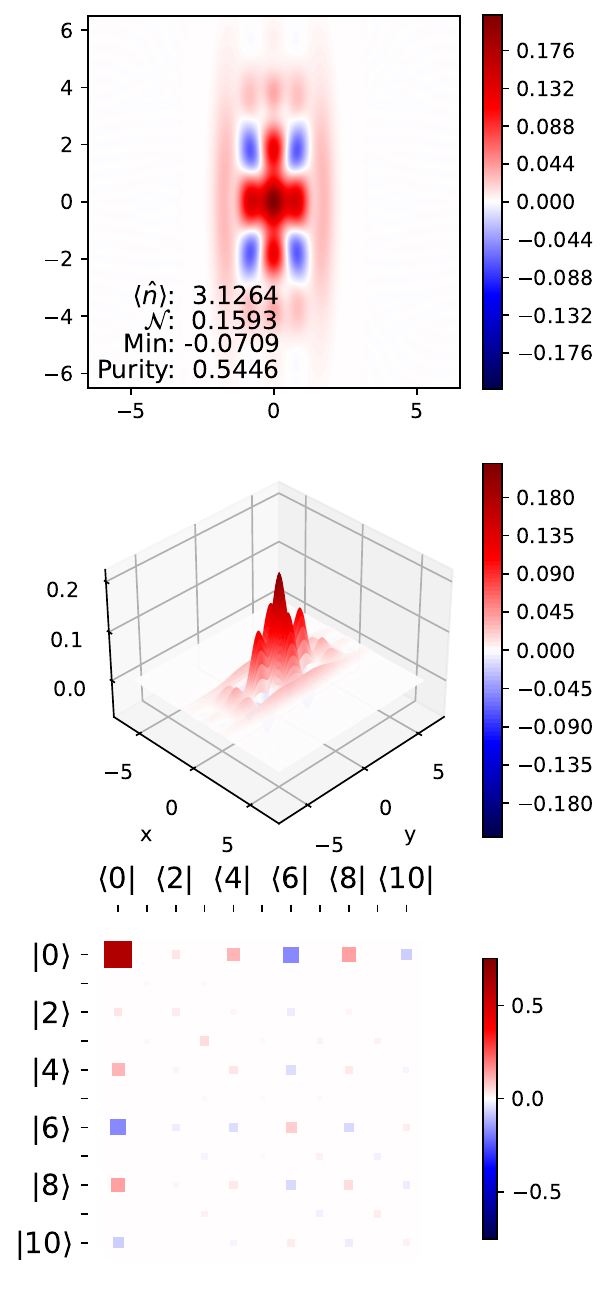}
    \caption{Wigner distribution (top and centre) and Hinton plot (bottom) for the state of maximum negative volume when driving a spin-1 emitter, using $r = 1.4679$,  $(T_{\rm s},T_{\rm c},t_{\rm c0})=(0.9318, 0.9707, 0.2226)\gamma^{-1}$. The Hinton plot has been truncated to the 10 photon state for visibility.}
    \label{figure:spin_one_squeezed_max_N_triple}
\end{figure}

In particular, we again use our variant of the FST-PSO algorithm to optimize the negative volume. This optimization finds the state shown in Fig.~\ref{figure:spin_one_squeezed_max_N_triple}, which is obtained for an initial squeezing parameter of $r = 1.4679$, and Gaussian temporal modes defined by $(T_{\rm s},T_{\rm c},t_{\rm c0})=(0.9318, 0.9707, 0.2226)\gamma^{-1}$. The  negative volume is $0.1576$. Whilst this is marginally greater than the maximum found in the pulsed squeezed driving of a spin-1/2 emitter, it does not show the same significant increase in negative volume as we found for higher-spin emitters in the pulsed coherent driving case. 

The present optimization is a significantly more resource- and time-intensive process due to the increased dimensionality of the system. Whilst we are confident this is the maximum negative volume that could be produced when driving a spin-1 emitter with a Gaussian temporal mode, we believe that higher negative volumes could be found by considering more general temporal modes.

\begin{figure}[ht]
    \centering
    \includegraphics[width=\linewidth]{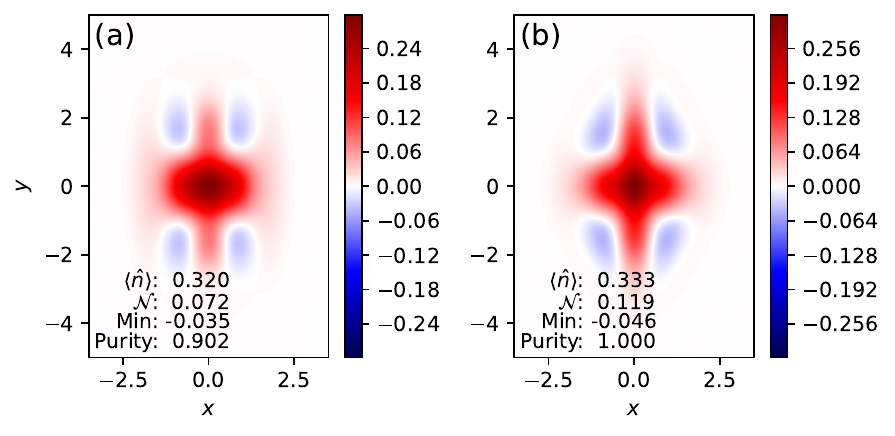}
    \caption{Wigner distributions for (a) the output temporal mode state when pulsing a squeezed state of light with squeezing parameter $r = 0.75$ through a spin-1 emitter, with Gaussian temporal modes defined by $(T_{\rm s},T_{\rm c},t_{\rm c0})=(0.93, 0.97, 0.22)\gamma^{-1}$, and (b) the state $N\left(\hat{I} - \ket{2}\bra{2}\right)\hat{S}(0.7)\ket{0}$, where $N$ is a normalisation factor. The two states have a fidelity of $0.9654$, indicating a high degree of similarity.}
    \label{figure:spin_one_exclusion_grid}
\end{figure}

Finally, it is very interesting to consider the Hinton plots of these states. Specifically, as shown in Fig.~\ref{figure:spin_one_squeezed_max_N_triple}, we observe an almost complete exclusion of a two-photon component in the temporal-mode state. In fact, for low input squeezing parameters, these states show considerable similarities with squeezed states that have simply had their two-photon component removed, i.e., states of the form
\begin{equation}
    \ket{\psi} = N\left(\hat{I} - \ket{2}\bra{2}\right)\hat{S}(r)\ket{0},
\end{equation}
where $N$ is a normalisation factor. This is illustrated in Fig.~\ref{figure:spin_one_exclusion_grid}, where the Wigner distribution of this state is compared with the distribution of a particular temporal mode state for small $r$.

\section{Correlation Function Analysis}\label{section:correlation}

A possible explanation for the two-photon exclusion shown in Section \ref{section:spinone} can be found by considering the first order correlation function for the outgoing field,
\begin{align} \label{equation:full_correlation}
    G^{(1)}(t,\,t') &= \langle \hat{a}_{out}^\dagger(t')\hat{a}_{out}(t)\rangle, 
\end{align}
where 
\begin{align}
\hat{a}_\text{out}(t) = \sqrt{\beta\gamma} \hat S_- + g_{\rm s}^*(t)\hat a + \hat{a}_\text{in}(t) .
\end{align}
After performing a spectral decomposition of $G(t, t')$ with respect to a set of orthogonal modes, such as the Hermite-Gaussian functions, we can determine the mean photon number of each mode \cite{input_output_two}. Importantly, the eigenmodes, $v_i(t)$, of the correlation function are another such set of modes and always contain the most occupied temporal mode, with each eigenvalue defining the photon occupation of the corresponding eigenmode. 

\begin{figure}[ht]
    \centering
    \includegraphics[width=\linewidth]{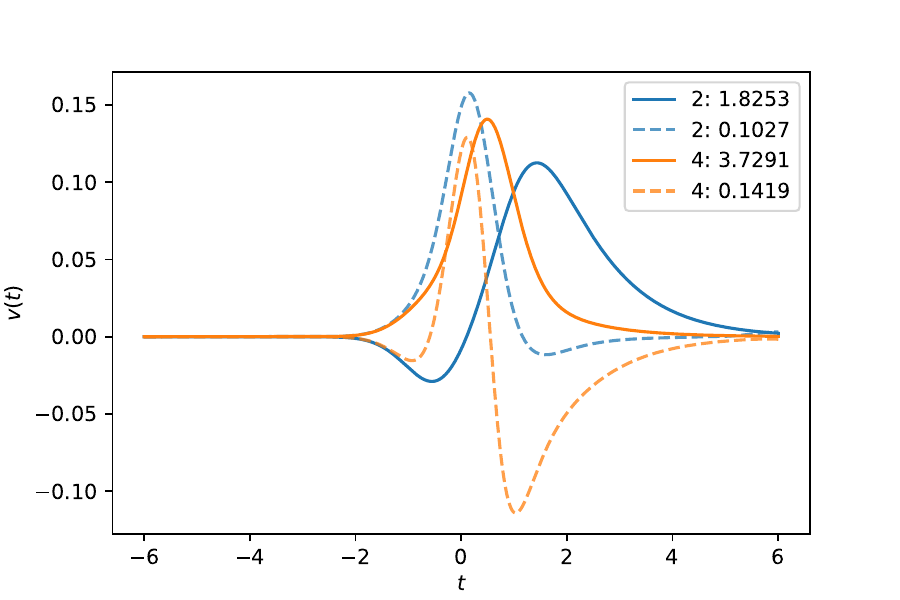}
    \caption{Dominant eigenmodes of the correlation function $G(t, t')$ of a spin-1 emitter driven by a two- or four-photon pulse in a Gaussian temporal mode with FWHM $1.6\gamma^{-1}$. Solid lines show the most occupied temporal mode, whilst dashed lines show the next most occupied orthogonal mode, with the legend showing the occupation of each mode.}
    \label{figure:two_four_photon_correlation}
\end{figure}

In order to investigate the observed two-photon exclusion, we examine the output-field correlation function for two- and four-photon pulses incident on a spin-1 emitter. Dominant eigenmodes are shown in  Fig.~\ref{figure:two_four_photon_correlation}, for an input pulse in a Gaussian temporal mode of FWHM $T_s = 1.6\gamma^{-1}$. Displayed are the two highest-occupation eigenmodes in each case.

If we compare the dominant temporal modes for each case, we see a stark difference between the two. For the four-photon driving, we see that over 90\% of the incident photon number is contained within a temporal mode that is approximately Gaussian. However, for the two-photon driving over 90\% of the incident photon number is contained within a starkly different mode, more closely resembling a first-order Hermite-Gaussian function.

\begin{figure}[ht]
    \centering
    \includegraphics[width=\linewidth]{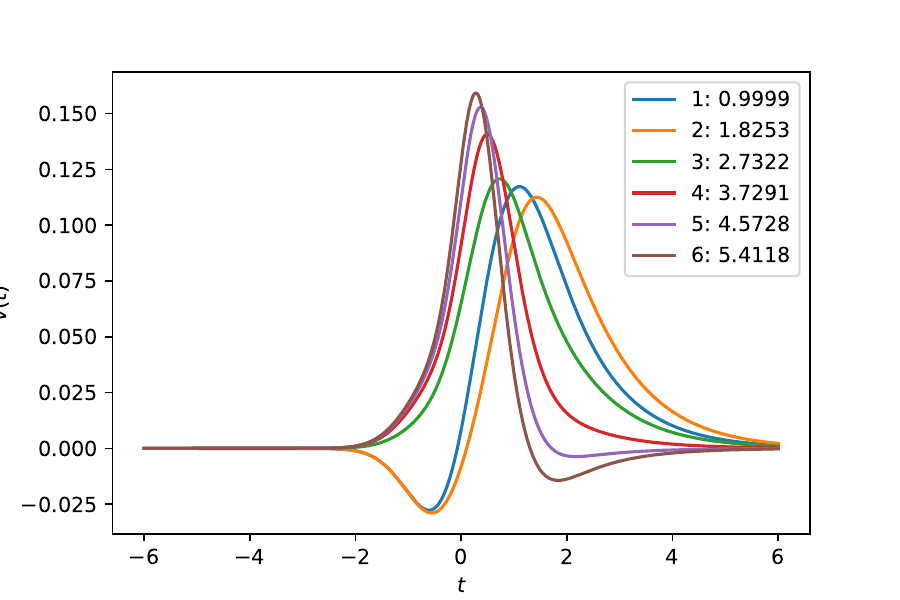}
    \caption{Dominant eigenmodes of the correlation function $G(t, t')$ of the outgoing field mode of a spin-1 emitter driven by a photon-number pulse in a Gaussian temporal mode with FWHM $1.6\gamma^{-1}$. The legend shows the photon number in the incident pulse, and the occupation of each eigenmode.}
    \label{figure:N_photon_correlation}
\end{figure}
If we now consider only the dominant temporal modes scattered from a spin-1 emitter driven with incident one- to six-photon pulses, displayed in Fig.~\ref{figure:N_photon_correlation}, which all contain over 90\% of the incident photon number, we see that the cases involving three or more photons display a similar, near-Gaussian shape, which greatly differs from the one- and two-photon-pulse cases.

We propose that this is a general property when driving spin-$S$ emitters with an $N$-photon pulse of light. Any photon number pulse up to $N = 2S$, i.e., any photon number pulse that could be fully absorbed by the emitter, has an output field dominated by a temporal mode that only weakly overlaps that for driving pulses with greater than $2S$ photons. 

As such, for the squeezed driving of a spin-1 system, the two-photon component of the squeezed state is emitted predominantly into a temporal mode that does not match the Gaussian mode we ``capture'', and thus the two-photon component is largely excluded. In contrast, the four- and above-photon components are contained predominantly in approximately Gaussian modes, and as such are mostly captured.

\section{Conclusion}\label{section:conclusion}
In summary, we have used a cascaded-systems model to numerically calculate the Wigner distributions of propagating temporal modes in the output field of a two-level emitter driven by pulses of coherent or squeezed light. We have shown that strong Wigner-negativity exists in these modes, far surpassing that generated in the case of driving by steady-state coherent or squeezed light. We have also found numerically the maximum Wigner-negative volumes that can be generated using this system, and confirmed the similarities of the generated states, in the coherent case, with displaced single-photon states, and, in the squeezed case, with squeezed Schr{\"o}dinger-cat states. 

Furthermore, we have demonstrated that this effect is not unique to two-level emitters. We have shown that pulsed coherent driving of spin-1/2 to spin-2 emitters can produce a variety of strongly Wigner-negative states, and that the pulsed squeezed driving of a spin-1 emitter can generate entirely new states of light. 
Preliminary investigations of the generated states have also shown high expectation values of GKP stabilizers, and thus strong similarities with GKP states. A deterministic source of optical GKP type states is also an exciting prospect, and thus further investigation is ongoing to determine the optimal GKP-like states that can be found with our setup. 

Another potential direction for investigation is that of investigating the quantum Fisher information associated with the system \cite{quantum_fisher_information}. Recently, similar works have investigated the information that outgoing field modes contain about a coherently-driven two-level emitter \cite{metrology_two_level}. We believe that, in our system, the outgoing field modes may contain significant information about the emitter, especially when reintroducing a nonzero detuning, and we are pursuing this line of investigation further.

{\em Note added}. During the completion of this manuscript, we became aware of related theoretical work examining temporal modes scattered from two-level emitters driven by coherent \cite{Lim2026} and squeezed \cite{Copie2026} light pulses. Lim {\em et al.}  \cite{Lim2026} consider optimization of the temporal-mode Wigner negativity using energetic cost functions and demonstrate efficient generation of a displaced single-photon state. Copie {\em et al.} \cite{Copie2026} also consider squeezed driving and demonstrate generation of approximate squeezed cat states, as in the present work, although for non-Gaussian input and output temporal modes.

\section*{Acknowledgments} 

The authors thank Miriam Leonhardt for help during the early stages of this work.

\bibliography{refs.bib}\label{section:refs}

\end{document}